\documentclass[twocolumn,amssymb, nobibnotes, floatfix, aps, prb]{revtex4-2}
\usepackage[utf8]{inputenc}
\usepackage[english]{babel}
\usepackage{amssymb}
\usepackage{amsmath}
\usepackage{graphicx}
\usepackage{color}
\usepackage{physics}
\usepackage{mathtools}
\renewcommand{\Re}{\operatorname{Re}}

\renewcommand{\vec}[1]{\boldsymbol{#1}}
\usepackage{hyperref}
\graphicspath{{Figs/}}
\begin{document}
\title{Flat bands, strains, and charge distribution in twisted-bilayer hBN}
\author{Niels R. Walet$^1$}
\email{Niels.Walet@manchester.ac.uk}
\homepage{http://bit.ly/nielswalet}
\author{Francisco Guinea$^{2}$}
\email{Francisco.Guinea@imdea.org}
%\homepage[]{Your web page}
%\thanks{}
\affiliation{$^1$Department of Physics and Astronomy, University of Manchester, Manchester, M13 9PY, UK}
\affiliation{$^2$Imdea Nanoscience, Faraday 9, 28015 Madrid, Spain}
\date{\today}

\begin{abstract}
We study the effect of twisting on bilayer graphene. The effect of lattice relaxation is included; we look at the electronic structure, piezo-electric charges and spontaneous polarisation. We show that the electronic structure without lattice relaxation shows a set of extremely flat in-gap states similar to Landau-levels, where the spacing scales with twist angle. With lattice relaxation we still have flat bands,
but now the spectrum becomes independent of twist angle for sufficiently small angles. We describe in detail the nature of the bands, and study appropriate continuum models, at the same time explaining the spectrum
We find that even though the spectra for both parallel an anti-parallel alignment are very similar, the spontaneous polarisation effects only
occur for parallel alignment. We argue that this suggests a large interlayer hopping between boron and nitrogen.
\end{abstract}

	\maketitle
	\section{Introduction}
	The field of twisted bilayer materials has literally exploded in the last few years after the discovery of 
	highly correlated phases in magic-angle twisted bilayer graphene (MATBG) \cite{Cao}.
	The superconducting and insulating phases seen in such materials as a function of doping suggests that interactions play a crucial role

	Many other materials have been studied, both theoretically and experimentally, including transition metal dicalchinides (TMDCs) \cite{Wu2018,Wu2019}, multilayer graphene systems such as twisted double bi-layers\cite{Chebrolu2019,Koshino2019,Pantaleon2020}, graphene stacks, and various forms of graphene twisted relative to hexagonal boron-nitride (hBN)\cite{Moon2014,San-Jose2014,Jung2017,Brown2018}, see also\cite{Cea2020,Lin2020,Mao2020,Shi2020}. Twisted hBN has also been proposed and studied\cite{Xian19,Zhao2020}, and is the subject of this work.
	
	There has been great recent interest in the electric properties of twisted-bilayer hBN \cite{woods2020chargepolarized,stern2020interfacial,yasuda2020stackingengineered,Zheng2020}, where spontaneous charge polarisation has been discovered for what is called "parallel" alignment, and none for the antiparallel one. Also, by mounting hBN on a conducting substrate we can look at the effect of an electric field. That leads to the question of the electronic structure of such materials: it is well known that flat bands occur in many such systems near the Fermi level, which drive most of the interacting physics since these are exquisitely sensitive to even  weak residual forces. This clearly deserves investigation.
	
	In MATBG such continuum models are usually based on what is now called the Bistritzer-McDonald  \cite{dos2007graphene,BM2011}, an in-layer continuum Dirac Hamiltonian with a very distinct 3-fold symmetry of the interlayer coupling. We have shown previously how we can derive a more detailed model from a tight-binding approach \cite{GuineaWalet2019,Walet_2019}, still keeping many of the simplicity of such a model. Other approaches are discussed in the literature, for instance in Ref.~\cite{Balents_2019,Carr_2019,carr2020electronic}.
	
	There is a n interesting question where we can find  flat bands, their nature and the continuum model description. With our toolbox we should be able to answer those questions, and we shall show that lattice relaxation has a surprising effect on the spectra, making an importnat part of it independent of twist angle.

	The relaxation of lattice gives rise to strain, and in piezo-electric materials this leads to charging in the areas of large strain. Since hBN is one of this nature, we could expect that charges are generated by the lattice relaxation. That leads to the question whether this is the dominant mechanism for charge generation, and whether it is responsible for the charge domains observed in experiment.
	
	The paper addresses these questions in order. We shall first look at lattice relaxation using an atomistic force model. We will then investigate the nature of the electronic states and especially the in-gap flat bands in twisted hBN. We shall look at the nature of these states in both rigid and relaxed hBN layers, and discuss continuum models that can be used to describe this. Finally we turn our attention to the nature of the charge domains.

	\section{Twists and relaxations}
	There are two models predominantly used for relaxation: one is the use 
	is a simple harmonic potential model, often linked to DFT calculations, as done by e.g., \cite{NamKoshino_2017,Enaldiev_2020,enaldiev2020piezoelectric,leconte2019relaxation,Tritsaris_2020}. Such models work surprisingly well, but lack some of the atomic detail for the smaller angles, which seems to lead to the occurrence of higher harmonics in the lattice deformation \cite{GuineaWalet2019}. The only practical way this atomic nature can be reinstated is by using classical potential models, the approach taken here. Of course, such an approach has it own limitations.
		\subsection{Potential model}
		We use a standard approach, using LAMMPS \cite{plimpton_fast_1995} to minimize the energy using a classical potential model to find atomic positions. One of the most approriate potential models for hBN seems to be the ``inter-layer potential" (ILP) from Refs.~\cite{leven_interlayer_2016, maaravi_interlayer_2017}. This has been bench-marked with the Tersoff in-layer potential \cite{Tersoff1988new,tersoff1989modeling,los2017extended}, which is what we shall adopt in this work. Of course, one should have a healthy scepticism as to the correctness of this in all details (see also below).
		
		Some large-twist-angle DFTB calculations for bilayer hBN systems have been performed by Zhao \emph{et al} \cite{zhao2019flat}. Another work along similar lines is that by Xian \cite{Xian19}. They do find flat bands, but the limitations of their
		computational techniques probably mean we can only use these results as indicative. The work by Javvaji \emph{et al} \cite{Javajji20} takes a similar approach to our work, but starts with a reduced model (but based on a tight-binding model like ours), and thus has some common elements: but there are also clear differences: the main one is that their continuum model is not correct for small twist angles (see discussion below).
		
%		\subsection{$0$ and $60^\circ$}
%		The same alignments can be found at $\theta \approx 0^\circ$, and near $60^\circ$ degree alignment. The unit cell used in our previous work, labelled by the two integers $(n,m)$ has a 3 times larger unit cell in area near $60^\circ$ ($m=1$) than near $0^\circ$ ($m=n-1$), and we find that the larger unit cells allows  for perfect alignment at intermediate angles. We believe that this is irrelevant for a discussion of the physics at small alignment angles, and we shall thus concentrate on this situation  
		\begin{figure}
			\includegraphics[width=0.75\columnwidth]{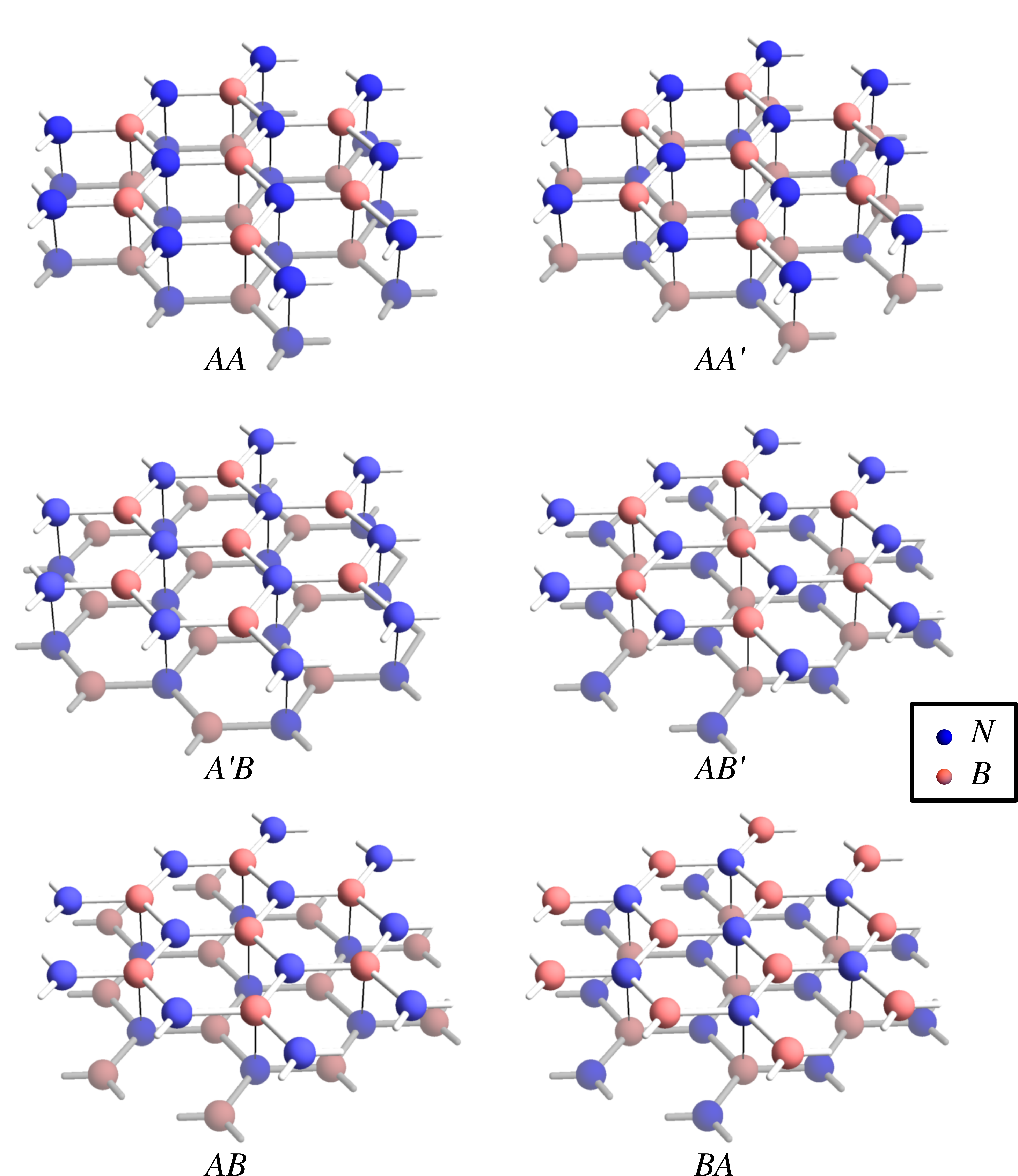}
			\caption{The five different alignments, labelled as in Ref.~\cite{RibeiroPeres11}, plus the top-bottom inversion $BA$. Vertical black lines denote aligned atoms.}
			\label{fig:alignment}
		\end{figure}
		
		\subsection{alignment}
		Since we have 5 potential alignments with complex energetics, see, e.g., Refs.~\cite{RibeiroPeres11,Constantinescu13} and Fig.~\ref{fig:alignment}, we will have to extend the analysis of our previous work. The main difference is that if we invert one of the layers, we change the main alignment, since we swap boron with nitride atoms: this corresponds to what is  called "anti-alignment" in the literature.
		\subsection{results} 
\begin{figure}
			\includegraphics[width=\columnwidth]{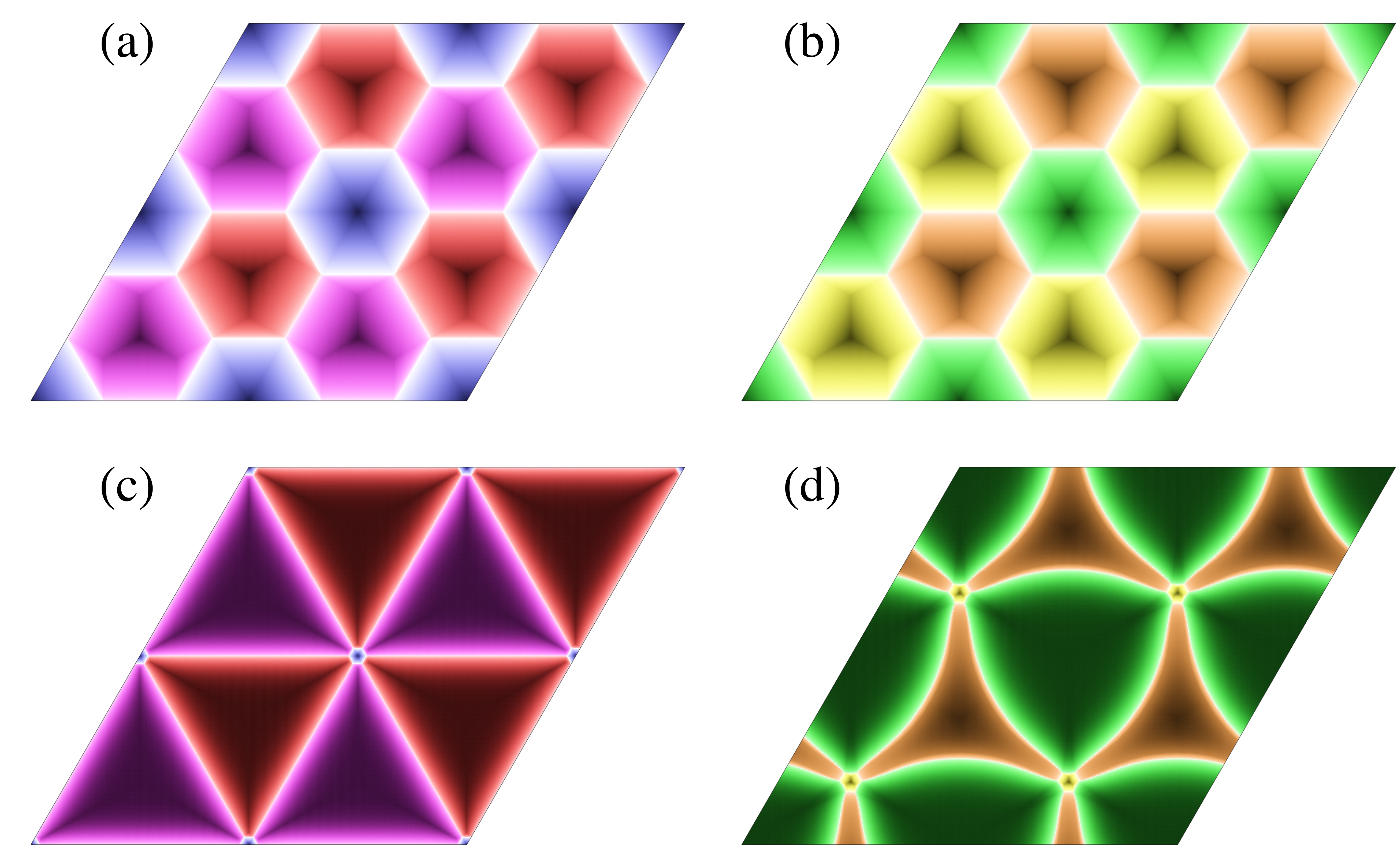}
			\caption{The dominant alignment structures (a,c) and (b,d) for a twist angle $\theta=0.33^\circ$, $L=43.15\,\text{nm}$. Each image shows four primitive cells. Images (a) and (c) are rigid, undeformed structures; (b) and (d) are the related associated relaxed structures. Each color shows a specific dominant alignment: Left column: Blue: $AA$ alignment (B above B, N above N); Purple: $AB$ alignment; red $BA$ (layer inversed) alignment.
			Right Column:
			Green: $AA^\prime$ (B in one layer above N in another); Orange/brown: $AB^\prime$ with aligned N; Yellow: $AB^\prime$ with aligned B. The darker the color, the stronger the alignment.}\label{fig:relaxation100x99}
		\end{figure}

		In Fig.~\ref{fig:relaxation100x99} we show the alignments, defined using an extension of the method in our previous work \cite{GuineaWalet2019}, see Appendix \ref{app:align}. We use colour saturation to show the quality of the dominant alignment, and a specific colour for each type of this dominant alignment. The twist angle is $0.33^\circ$. 
		
		As we can see in the figure, we find a substantial reorganisation of alignment in both cases, with a very different pattern for aligned or anti-aligned hBN. Such patterns will induce an inhomogeneous strain in the hBN, and since the material is piezo-electric, will also induce charge (or in other words, the change in the in-layer hopping parameters, although small, will induce a charge). We have evaluated this in a "semi-continuous" way. The strain tensor is the Lagrangian finite strain tensor, $E=\frac{1}{2}(F_S^T F_S-I)$, where $F_S$ is evaluated using the method for discrete hexagonal lattices from Ref.~\cite{pozrikidis2008mechanics}; we then turn this into a piezo-electric charge using the method in Ref.~\cite{rostami2018piezoelectricity}. Derivatives are required in that method are replaced with a finite difference on the hBN lattice sites. 
		
		\begin{figure}
			\includegraphics[width=\columnwidth]{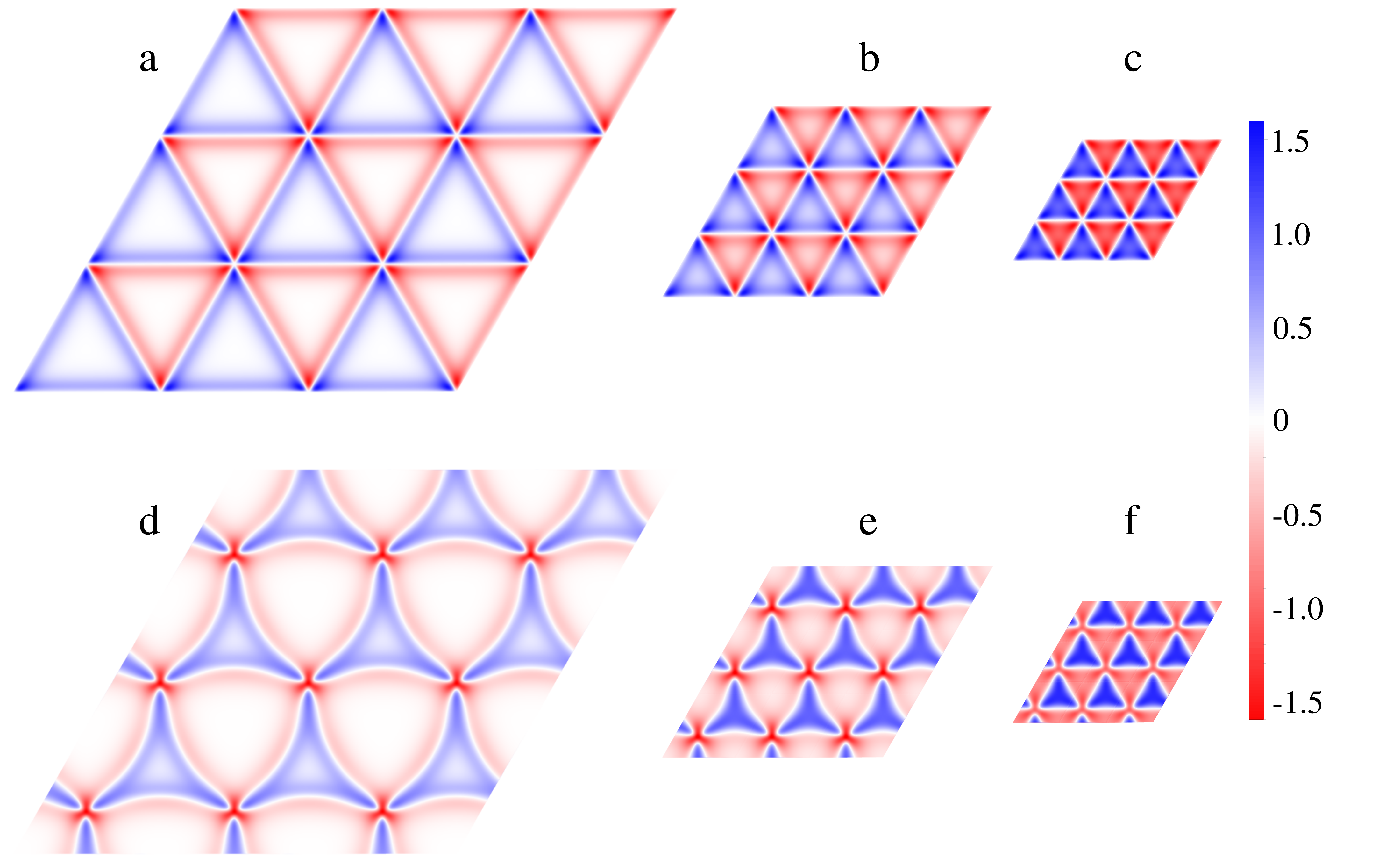}
			\caption{The induced piezo-electric charge in a single layer after relaxation of a hBN bilayer. a-c: aligned at angles a) $0.33^\circ$ b) $0.67^\circ$ and c) $1.05^\circ$. d-e: anti-aligned at angles d) $0.33^\circ$ e) $0.67^\circ$ and f) $1.05^\circ$. The electron density $n$ (scale on the right) is given in units of $10^{12}\,\text{cm}^{-2}$. All images are drawn to the same scale.}\label{fig:piezo}
		\end{figure}
		
We notice that charge density concentrates around the channels where the alignment changes, but also that the charge density seems to decrease as we decrease the angle; somewhat the opposite from what one would naively expect. It also seems to be different than what the experimental data says\cite{woods2020chargepolarized,stern2020interfacial,yasuda2020stackingengineered,Zheng2020}.
		
	\section{Electronic structure}
	Even though the piezo-electric effect discussed above is in essence due to electronic structure, but we shall first concentrate on the electronic hopping in the bilayer system.
	As discussed before, there are a few papers that discuss the spectrum of hBN.
	Core for the work here is the paper by Ribeiro and Peres \cite{RibeiroPeres11}, the first to derive a tight binding model for a bilayer--the ingredient crucial to our work.
	
	They start form a rather simple DFT calculation of two infinite bilayers aligned in the 5 positions show in Fig.~\ref{fig:alignment}. Weaknesses of their DFT inputs are a substantial mismatch between the calculated and actual layers distance (the calculated one in their work is substantially larger than the accepted value of $3.33\,\text{\AA}$), and what seems to be a gross underestimate of the gap--they find values around $4\,\text{eV}$, and a GW calculation may be closer to $8\,\text{eV}$ \cite{pedersen2015intraband,WirtzMR06,AttaccaliteBMRW11,HueserOT13}. For some reason unclear to us, rather than fitting a single tight binding model to all alignments, they fit a different tight binding model to each alignment.
	
	Since the idea to fit a tight binding model is reasonable we use a single more complete version of such a mode and see what physics we can describe: We shall make use of an exponential parametrisation of the interlayer hopping parameters, and will ignore the nearest neighbour in-layer one. Thus our interlayer hopping parameters will be assumed to take the simple, and potentially still too naive, form
	\begin{equation}
	t_{XY}(r)=t_{XY}\exp(-\alpha(r-d))\,.\label{eq:hopping}
	\end{equation}
	We shall use $d=3.33\,\text{\AA}$ and $\alpha=4.4\,\text{\AA}^{-1}$.\footnote{We will look at other choices below.} We shall not fit these parameters to DFT, but instead use a reasonable guess based on other results. This makes sense, since, as we shall argue below, the qualitative results do not depend on these values, and it is well know that DFT struggles for insulating systems. 
	
	With these additional hoppings, if we truncate the interlayer coupling to atoms either placed directly above each other and their nearest neighbours, we find  Hamiltonian matrices that are a slight generalisation of those in Ref.~\cite{RibeiroPeres11}, see \eqref{eq:tb} for  detailed expressions. These depends on
	the gap  $\Delta$, the in-layer hopping $t$.
	Unlike in Ref.~\cite{RibeiroPeres11} we assume that these parameters are the same for all alignments.
	\begin{figure}[t]
		\includegraphics[width= \columnwidth]{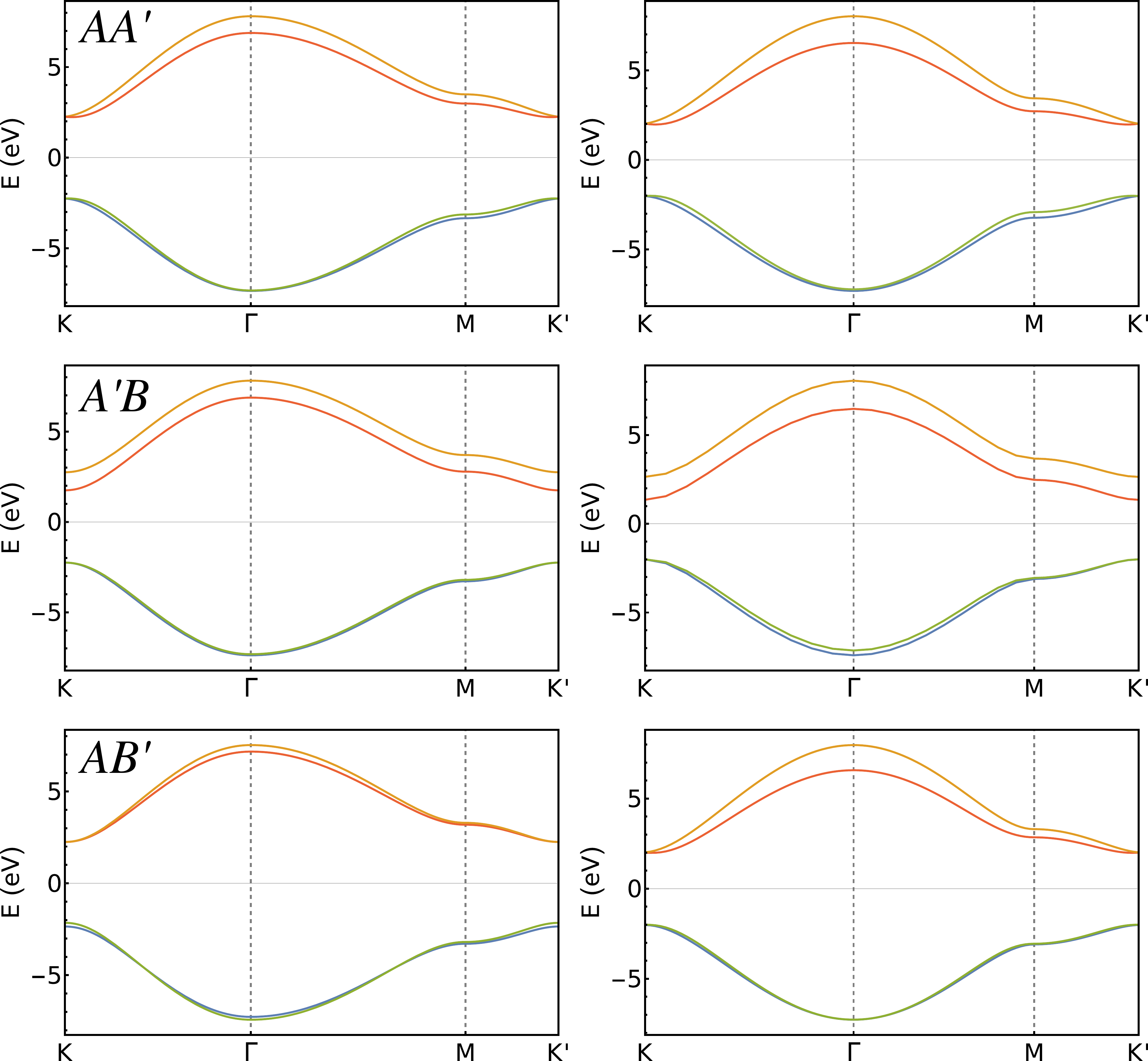}
		\caption{Right column: spectra for the naive tight binding model for three different alignments as labelled. We use $\Delta=4.5\,\text{eV}$, $t=2.33\,\text{eV}$, $t_{BB}=0.5\,\text{eV}$, $t_{NN}=0.1\,\text{eV}$ and $t_{NB}=0.2\,\text{eV}$. The decay is parametrised by $x=0.25$. We also show, in the right column the results of the tight-binding model including longer-range  interlayer hoppings.}\label{fig:spectra1}
	\end{figure}
	
	With the parameters as shown in Fig.~\ref{fig:spectra1} we get a reasonable representation of the spectra and gap as compared to DFT. Since we are describing a system with smaller $d$  and larger gap than in Ref.~\cite{RibeiroPeres11}, in the end we use the parameters
	$\Delta=8\,\text{eV}$, $t=2.33\,\text{eV}$, $t_{BB}=0.7\,\text{eV}$, $t_{NN}=0.15\,\text{eV}$ and $t_{NB}=0.3\,\text{eV}$ as probably more representative (we shall argue below that experimental data suggest that $t_{NB}$ may well be larger). We ignore next-nearest neighbour in-layer couplings; as shown in Ref.~\cite{chegel2016engineering}, their effect is small, and next-nearest neighbour hoppings would require us to determine two additional parameters for the calculations.
	
	We can now use these parameters to try and find the spectra of twisted hBN layers.
	For computational efficiency, we shall initially study an angle of $1.05^\circ$--full tight binding models are rather expensive for smaller angles, but we shall investigate an alternative approach in a later section.
	
	\subsection{Detailed comparison}
\begin{figure}
		\includegraphics[width=\columnwidth]{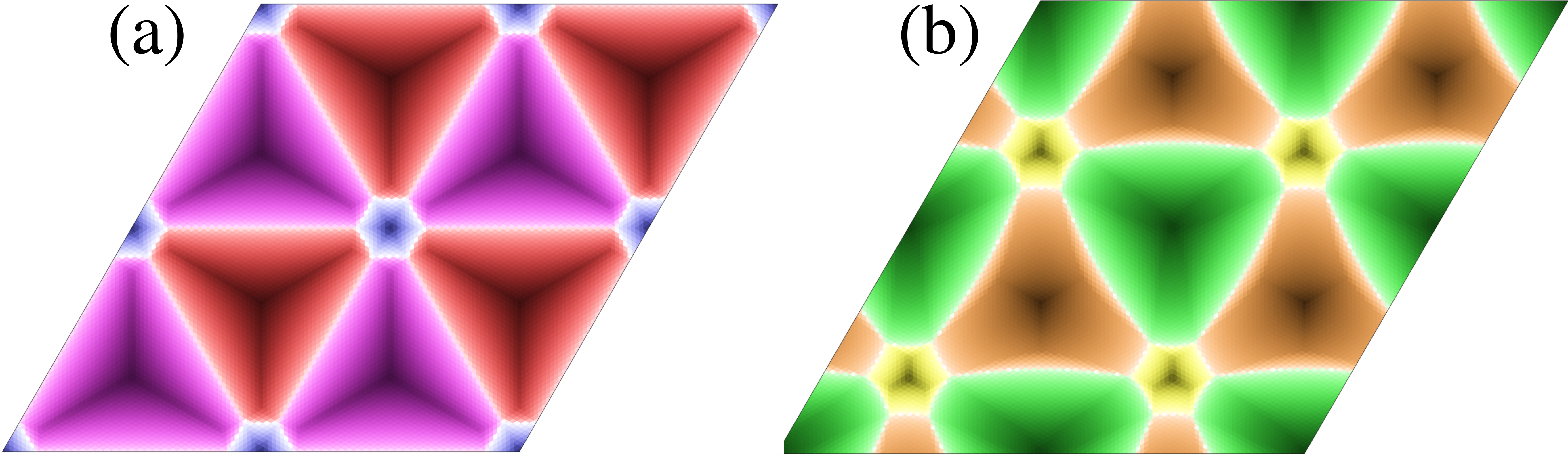}
		\caption{Relaxation for $\theta=1.05^\circ$. See caption of Fig.~\ref{fig:relaxation100x99} for details. Each image shows four primitive cells.}\label{fig:32x31_1}
	\end{figure}

\begin{figure}
    	\includegraphics[width=\columnwidth]{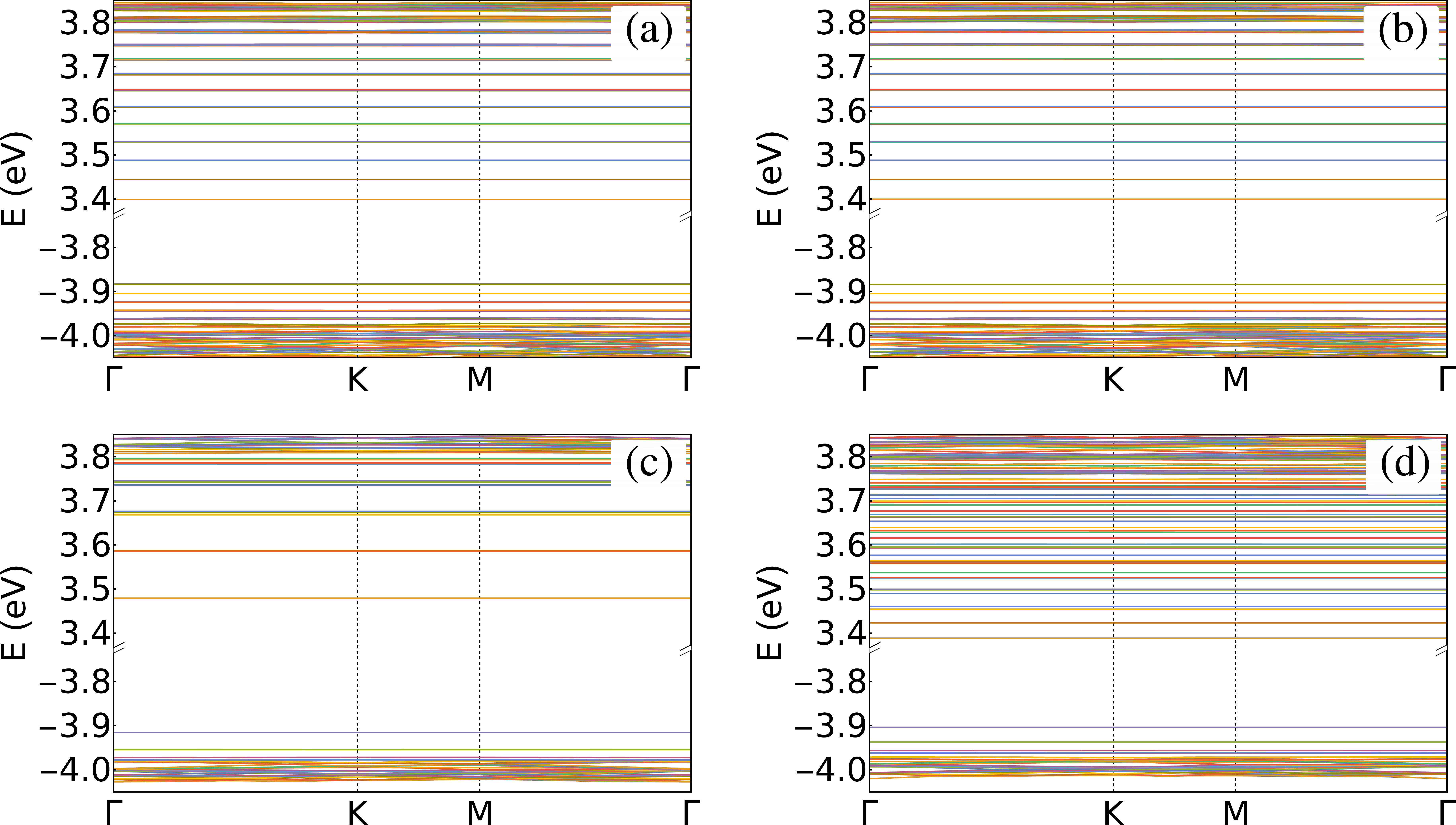}
    	\caption{Tight-binding spectra for $\theta=1.05^\circ$. a) unrelaxed  and c) relaxed lattice corresponding to a) in  Fig.~\ref{fig:32x31_1}. b) unrelaxed  and d) relaxed lattice corresponding to b) in  Fig.~\ref{fig:32x31_1}.}\label{fig:32x31_allsp}
    \end{figure}

	We start with a  detailed tight-binding calculation for a twist angle of $1.05^\circ$, a cell spacing of $13.6\,\text{nm}$. The alignment for the two relaxed structures considered here is show in Fig.~\ref{fig:32x31_1}.
    We have calculated the spectra for both relaxed and unrelaxed structures of this nature, see Fig.~\ref{fig:32x31_allsp}. We note that the spectra for the unrelaxed structures are identical--but the Hamiltonians are completely different, and thus we need to identify a mechanism that gives rise to this! The spectra for these two cases seem similar to Landau levels, which is also reflected in the increasing degeneracy: 2 for the states deepest inside the gap, then $4,6, 8,\ldots$  as the energy increases. As we shall show below, their origin is different from the argument made for twisted bilayer graphene in Ref.~\cite{liu_pseudo_2019}.
    
    For the relaxed positions we see rather different spectra, but in all cases we have just a simple valley degeneracy at the edge of the gap. In all cases the spectra are extremely flat: the bandwidth of each state is only a fraction  of an meV until we reach the quasi-continuum at the gap energy.
    
\begin{figure}
    	\includegraphics[width=\columnwidth]{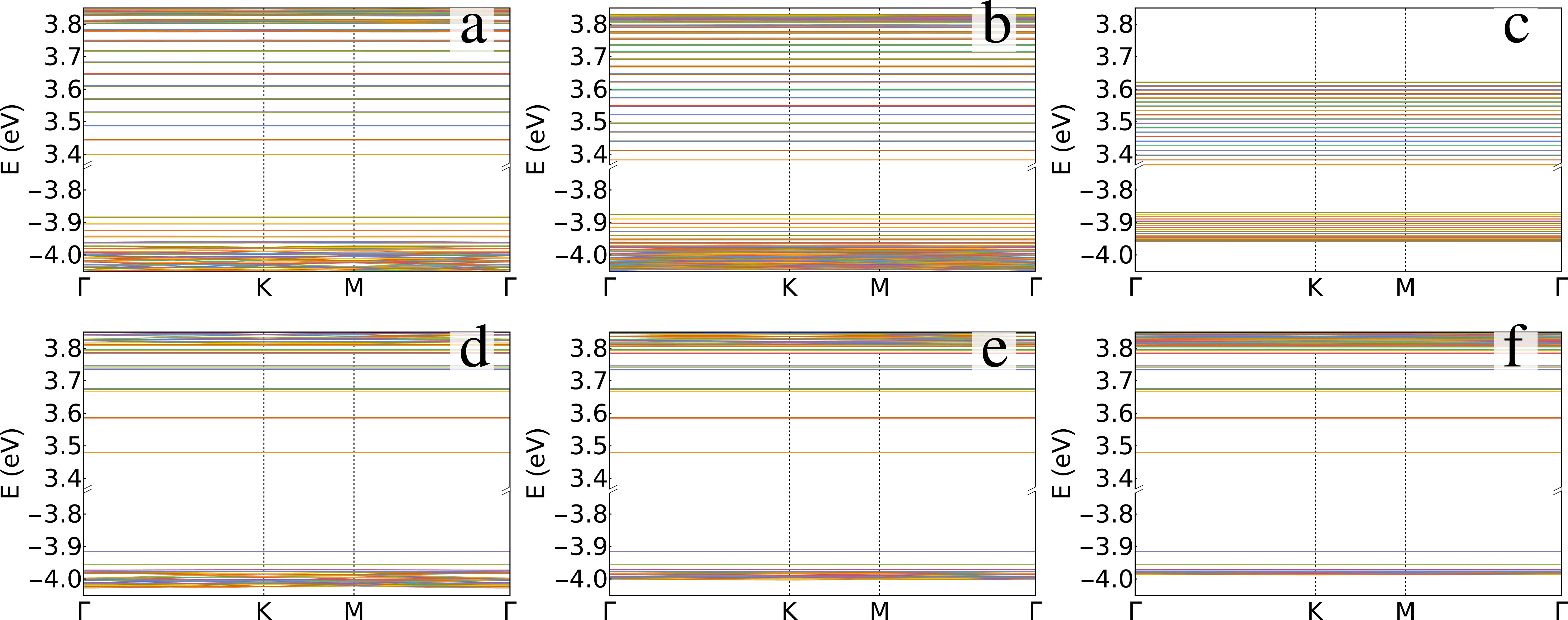}
    	\caption{Tight-binding spectra for (a,d) $\theta=1.05^\circ$,  (b,e) $\theta=0.67^\circ$, (c,f) $\theta=0.33^\circ$. This is for the relaxed lattice (c) in \ref{fig:relaxation100x99}. (a,b,c) no relaxation; (d,e,f): relaxed lattice. The sharp cut-off at top and bottom of the spectrum for the smallest angles is an artifact of our numerical approach due to the calculation of a finite number of eigenvalues.}\label{fig:100x99a}
    \end{figure}

There are some very intriguing features in spectra as we change twist angle: As can be seen in Fig.~\ref{fig:100x99a}, when there is no relaxation we  see an equally spaced set of levels that also show a typical two-dimensional harmonic oscillator degeneracy (doubled due the valley degeneracy), where the spacing decreases with an increasing moir\'e wave length.
On relaxation a few flat bands remain, with the same 2-4-6-\ldots degeneracy, but the most surprising result is that the in-gap states now appear to be independent of twist angle: there energies are so similar that  we had to check twice that we had actually used the right images!

We conclude that we will have to find an explanation for two different phenomena: the occurrence for in-gap flat bands (flat to within a fraction of an meV), which show a harmonic oscillator type spacing for lattices that do not relax at the interface, with the spacing decreasing as the twist angle decreases, and the occurrence of twist-angle independent flat bands if we relax the lattice at sufficiently small angles. Both of these should be described by a type of continuum model. The first case by a Bistritzer-McDonald like model discussed in the next section, and the second by an alternative approach, probably in real space.

	\section{Continuum projection}  

\begin{figure}
\includegraphics[width=\columnwidth]{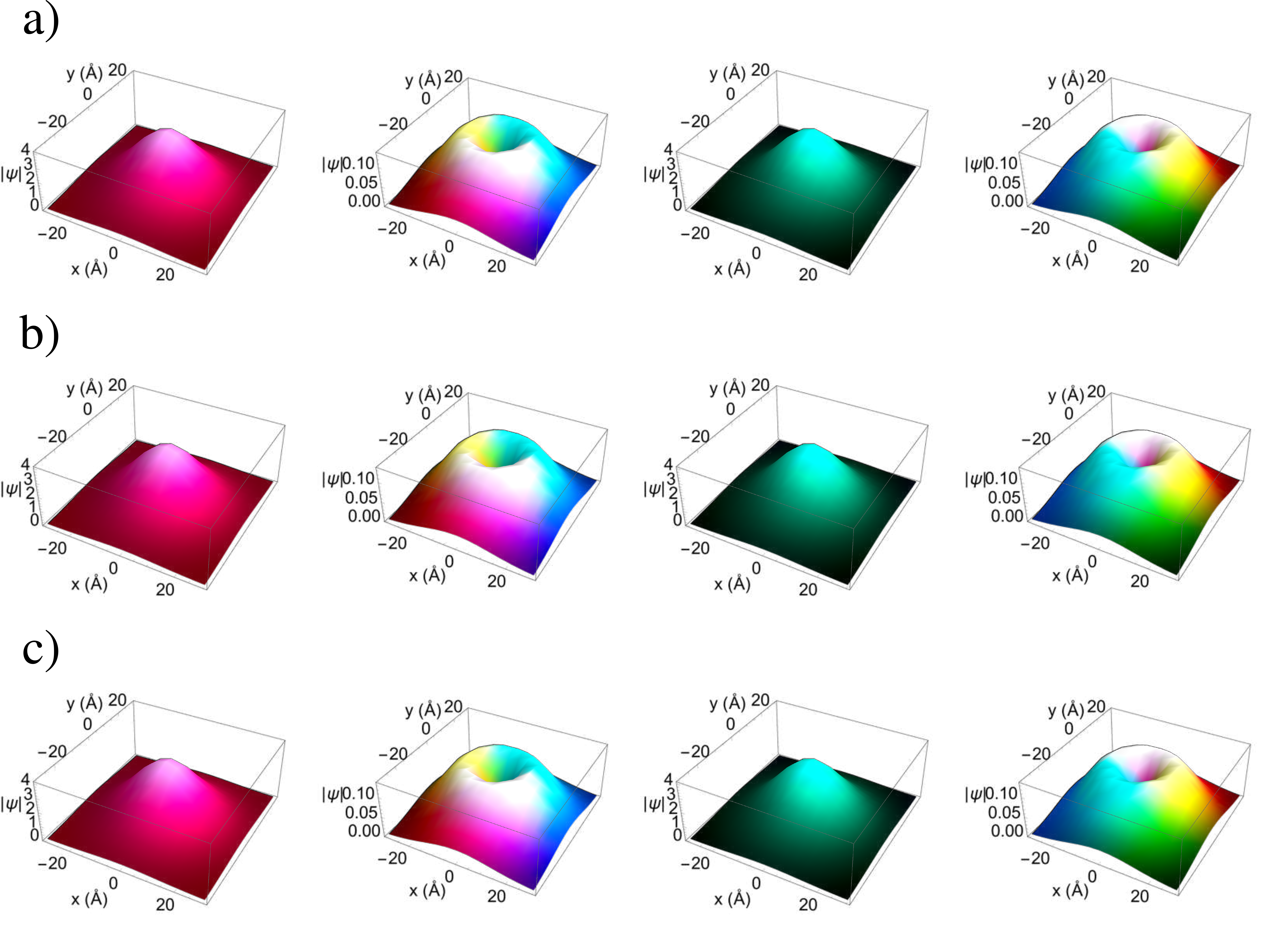}
\caption{the four components of the real space wave function multiplied with the phase $\exp(i(\vec k-\vec K_i)\cdot \vec r)$ for a) $\vec k=\vec K_1$, b) $\vec k=\vec K_2$ and c) $\vec k=\vec K_1/5+\vec K_2/3$. The four columns are layer 1 N (A) sites, layer 1 B sites, and the same for layer 2.
 The hue of the colouring shows the phase of the wave function. Thus the first column is real and positive, the third column real and negative. The B site wave functions show an (almost) uniform phase change of $2\pi$ as function of the polar angle around the origin.\label{fig:wfs}}
\end{figure}

	In order to understand the flatness of the bands, we first plot some wave functions from a tight binding calculation, and see that these indeed look like 2D harmonic-oscillator states shifted by a value proportional to the momentum. To get an analytical handle on this, we turn to a   continuum projection, using a ``generalised Bistritzer-MacDonald model" \cite{dos2007graphene,BM2011}. We follow the approach set out in \cite{GuineaWalet2019}. This established technology is known to generate continuum models that completely reproduce the full tight-binding calculations near the Fermi energy for graphene; we just need to check the equivalent result for hBN. 
	
		\begin{figure}
		\includegraphics[width=\columnwidth]{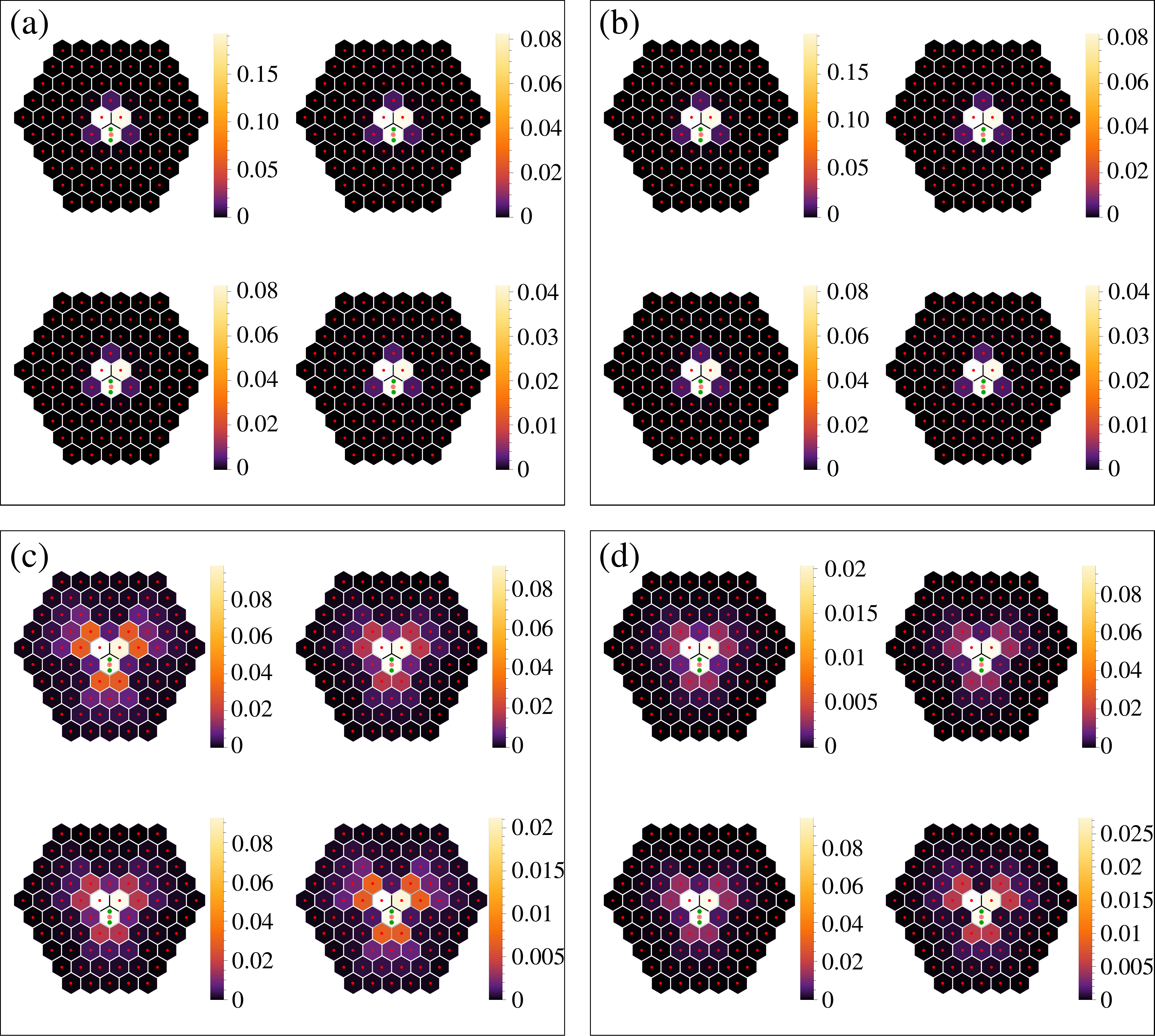}
		\caption{Projection of the interaction onto a continuum model as described in \cite{GuineaWalet2019}. Each hexagon is one interaction matrix element for a given momentum transfer. Zero momentum transfer is denoted by the pink dot, and the three dominant matrix elements for the unrelaxed lattice are exactly those for the BM model. In each case the left top figure of a pair shows the $AA$ projection, and the right top one the $AB$ case.
			The lower row show the $BA$ and $BB$ case, respectively. Unlike in graphene, the $AA$ and $BB$ case are not identical. The a-d labels are as in previous figures.}\label{fig:32x31_project}
	\end{figure}
	
	As we notice in Fig.~\ref{fig:32x31_project}, for undeformed lattices the projections of aligned and anti-aligned layers are identical. We have checked the Hamiltonians are rather different, but the results here are no surprise due to
 the similarity of the tight-binding spectra calculated earlier. Upon relaxation, we see that higher harmonics, corresponding to a larger superlattice momentum transfer, start playing a role.
 	
	We have investigated the source of the flat bands using this model--we shall concentrate here on the larger spacing at the positive side of the gap, but a similar analysis applies at the other end.
	For the case of the flat bands we have the classic  Bistritzer-MacDonald model with an extra gap added to the in-layer Hamiltonian,
\begin{align}
	H&=\begin{pmatrix} h(\vec K_1,\theta/2)
	 & U(\vec r)\\
	U^\dagger(\vec r ) & h(\vec K_2,-\theta/2)
	\end{pmatrix}\,,
	\\
	h(\vec k,\vec \theta)&=
	\vec {h}_\text{l}(-i\vec \nabla-\vec k,\theta)\cdot\vec \sigma+\frac{\Delta}{2} \sigma_3\,.
	\end{align}
	Unlike for the case of graphene, we find little benefit using the full in-layer tight-binding dispersion for $\vec{h}_\text{l}$ rather than the simpler linear expansion $\vec h_\text{l}(\vec k)=\hbar v_\text{F} \vec k$.
	
	In its simplest form the matrix $U$ takes the form
	\begin{equation}
	\begin{pmatrix}
	u_{AA}(\vec r) & u_{AB}(\vec r)\\
	u_{AB}^\dagger(\vec r) & u_{BB}(\vec r)
	\end{pmatrix}
	\end{equation}
	where \
	\begin{align}u_{AA}(\vec r)&=u_{0\text{BB}} g(\vec r),\\ 
	u_{BB}(\vec r)&=u_{0\text{NN}} g(\vec r),\\ 
	u_{AB}(\vec r)&=u_{0\text{BN}} g'(\vec r),
	\end{align}
	 with
	\begin{align}g(\vec{r})&=(1+e^{-i \vec G_1 \cdot \vec r}+e^{i \vec G_2 \cdot \vec r}),\\ 
	g'(\vec{r})&=(1+e^{-i2\pi/3}e^{-i \vec G_1 \cdot \vec r}+e^{i2\pi/3}e^{i \vec G_2 \cdot \vec r}).
	\end{align}
	
	\subsection{In gap states}
	
    We first look numerically which parameters are most relevant; we find that
	the energies are largely insensitive to the value of  $u_{0BN}$ and $u_{0NN}$, and the wave functions are dominantly located on the Boron  sites (there is a small component on the $\text{N}$ sites); they look very much like two (discrete) Gaussians centred on the momentum $\vec k-\vec K_i$; the signs of the Gaussians are opposite for the two layers. 
	We find that in this case the effect of replacing  $\vec {h}_\text{l}$ by its linear expansion is small, and we will thus work with the latter.

	The result we see brings to mind the analysis of Ref.~\cite{liu_pseudo_2019}, even though that work is for a different problem, and seems to ignore the mismatch between the two Dirac points, which is crucial for a cancellation of the gauge fields, see below.
	We look at the Hamiltonian in coordinate space, where the in-layer
	potential is expanded about the two $K$-points, following the standard "Bistritzer-MacDonald" continuum model \cite{BM2011}, with the addition of a gap. In order to simplify the analysis, we define the wave-function with a momentum translation to the relevant $\vec K$ point by writing
	\begin{align}
	\psi_{\vec{k}}(\vec  r)=
	\biggl(&e^{i \vec K_1\cdot \vec r}\psi_{B1\vec{k}}(\vec r),
	 e^{i \vec K_1\cdot \vec r}\psi_{N1\vec{k}}(\vec r),\nonumber \\
	 &e^{i \vec K_2\cdot \vec r}\psi_{B2\vec{k}}(\vec r),
	 e^{i \vec K_2\cdot \vec r}\psi_{N2\vec{k}}(\vec r)\biggr)\,.
	\end{align}
	\begin{widetext}
	We also permute rows and columns, so that the positive gap appears in the upper left-hand block,
\begin{equation}	
H=	\begin{pmatrix}
 {\Delta }/{2} & u_{aa} g(r) e^{i \delta \vec K \cdot\vec r} &  v_F p_- & u_{ab} {g'}(r) e^{i \delta \vec K \cdot\vec r} \\
 u_{aa} e^{-i \delta \vec K \cdot\vec r} g(r)^* & {\Delta }/{2} & u_{ab} {g'}(r) e^{-i \delta \vec K \cdot\vec r} &  v_F p_-  \\
  v_F p_+ & u_{ab} e^{i \delta \vec K \cdot\vec r} {g'}(r)^* & -{\Delta }/{2} & u_{bb} g(r) e^{i \delta \vec K \cdot\vec r} \\
 u_{ab} e^{-i \delta \vec K \cdot\vec r} {g'}(r)^* &  v_F p_+ & u_{bb} e^{-i \delta \vec K \cdot\vec r} g(r)^* & -{\Delta }/{2} \\
\end{pmatrix},
\end{equation}
where $\delta K$ is the difference in momenta between the two $K$ points,
$\delta \vec K=\vec K_1-\vec K_2$.   
We use the notation $\vec p$ to denote the momentum operator, with $p_\pm=p_1\pm i p_2$.
We assume we are looking at an in-gap eigenvalue just below the top of the gap, $E=\Delta/2-\epsilon$, with $\epsilon\ll\Delta$. We solve for the lower two components to eliminate the B wave functions, and find to first order in $1/\Delta$ [strictly speaking, we expand in terms of all five of the small scales $v_F\langle p\rangle /\Delta$, $u_i/\Delta$ and $\epsilon/\Delta$]:
\begin{equation}
H_0=	\begin{pmatrix}
 {\Delta }/{2} & 0\\
 0 & {\Delta }/{2} 
\end{pmatrix},
\end{equation}
%which is cancelled by re-expressing the eigenvalue in terms of $\epsilon$
and 
\begin{equation}
H_1=	\frac{1}{\Delta} \begin{pmatrix}
 -{u_{ab}^2 \left| {g'}(r)\right| ^2+ v_F^2 p^2} & {e^{i \delta \vec K\cdot\vec r} \left(\Delta u_{aa}g(r)-2 u_{ab}  v_F  \Re(p_+ g'(r))\right)} \\
 {e^{-i \delta \vec K\cdot\vec r} \left(\Delta u_{aa} g(r)^*-2 u_{ab}  v_F \Re(p_+ g'(r)))\right)} & -{u_{ab}^2 \left| {g'}(r)\right| ^2+v_F^2 p^2}
\end{pmatrix}\,.
\end{equation}
\end{widetext}
Intriguingly enough, we see that the kinetic energy actually only appears at first order. So where does the harmonic oscillator potential we hope to see appear?
In order to see that, we must re-express our results in the basis where the Hamiltonian $H_0$ is diagonal. We suffer from the problem that the basis now depends on $r$, and thus the kinetic energy acts non-trivially on this.

Let us first analyse what happens if we ignore the terms proportional $u_{ab}$--numerically we see these are unimportant relative to the lowest order potential due the fact that $u_{ab}\ll \Delta$, and performing the full analysis just hides some of the underlying simplicity. We diagonalise the Hamiltonian to first non-vanishing order, which gives two contributions: one due to the matrix diagonalisation transformation
\begin{equation}
H_{1}'=\begin{pmatrix}
  v_F^2/\Delta\, p^2  -u_{aa} \left| g (r)\right|& 0 \\
 0 &   v_F^2/\Delta\, p^2 +u_{aa} \left| g (r)\right|
\end{pmatrix}\,,
\end{equation}
and a second due to the fact that the momentum operator in $H_1$ acts non-trivially on the transformation matrix, which term can be written as
\begin{equation}
T=\frac{1}{\sqrt{2}}
\begin{pmatrix}
e^{i(\delta \vec K\cdot \vec r+\phi_g(r))}&1\\
-e^{i(\delta \vec K\cdot \vec r+\phi_g(r))}&-1
\end{pmatrix},
\end{equation}
with
\begin{equation}
e^{\phi_g(r))}=\sqrt{g(r)}/\sqrt{g(r)^*}.
\end{equation}
The derivative of the full phase can now be found as
\begin{equation}
\vec \nabla e^{i(\delta \vec K\cdot \vec r+\phi_g(r))}=
(\delta \vec K-\vec{G}_1e^{-i \vec G_1\cdot \vec r}+\vec{G}_2e^{i \vec G_2\cdot \vec r} ) e^{i(\delta \vec K\cdot \vec r+\phi_g(r))}.
\end{equation}
Since $\delta\vec K=\vec{G}_1-\vec{G}_2$, this quantity vanishes for small $r$, and thus we can safely ignore the resulting vector potential near the origin, where all the wave functions are located.

If we now look at the top entry of the  Hamiltonian, which describes the in-gap modes, we find an effective Hamiltonian
\begin{equation}
 H_\text{eff}=\frac{ v_F^2}{\Delta} p^2-u_{aa} \left| g (r)\right|\,.
\end{equation}
If we now expand $|g(r)|$, we find that we can write 
\begin{equation}
\left| g (r)\right|=3-\frac{4  \left(\pi ^2 \left(x^2+y^2\right)\right)}{3 L_s^2}\,.
\end{equation}
Thus all together, we have a harmonic oscillator
\begin{equation}
H_\text{eff}=-3u_{aa}+
\frac{v_F^2}{\Delta}
p^2 -\frac{4\pi u_{aa}}{3L_s^2} r^2\,
\end{equation}
Thus the spacing of the levels is 
\begin{equation}
\hbar\omega=\sqrt{4\frac{\hbar^2 v_F^2}{\Delta}\frac{4\pi u_{aa}}{3L_s^2}}
=\frac{4\hbar v_F}{L_s} \sqrt{\frac{\pi u_{aa}}{3\Delta}},
\end{equation}
with a lowest energy of $-3u_{aa}+\hbar\omega$. This agrees well with the results shown in Figs.~\ref{fig:100x99a}a--c.

\subsection{relaxation}
We have already seen in Fig.~\ref{fig:100x99a}d--f that for the relaxed latticed the in-gap spectra are roughly independent of twist angle. Since the Wannier functions corresponding to these bands are still like Gaussians, what we must have that the real-space continuum model describing these states  is essentially the same. 

  	 \begin{figure*}
  		\includegraphics[width=\textwidth]{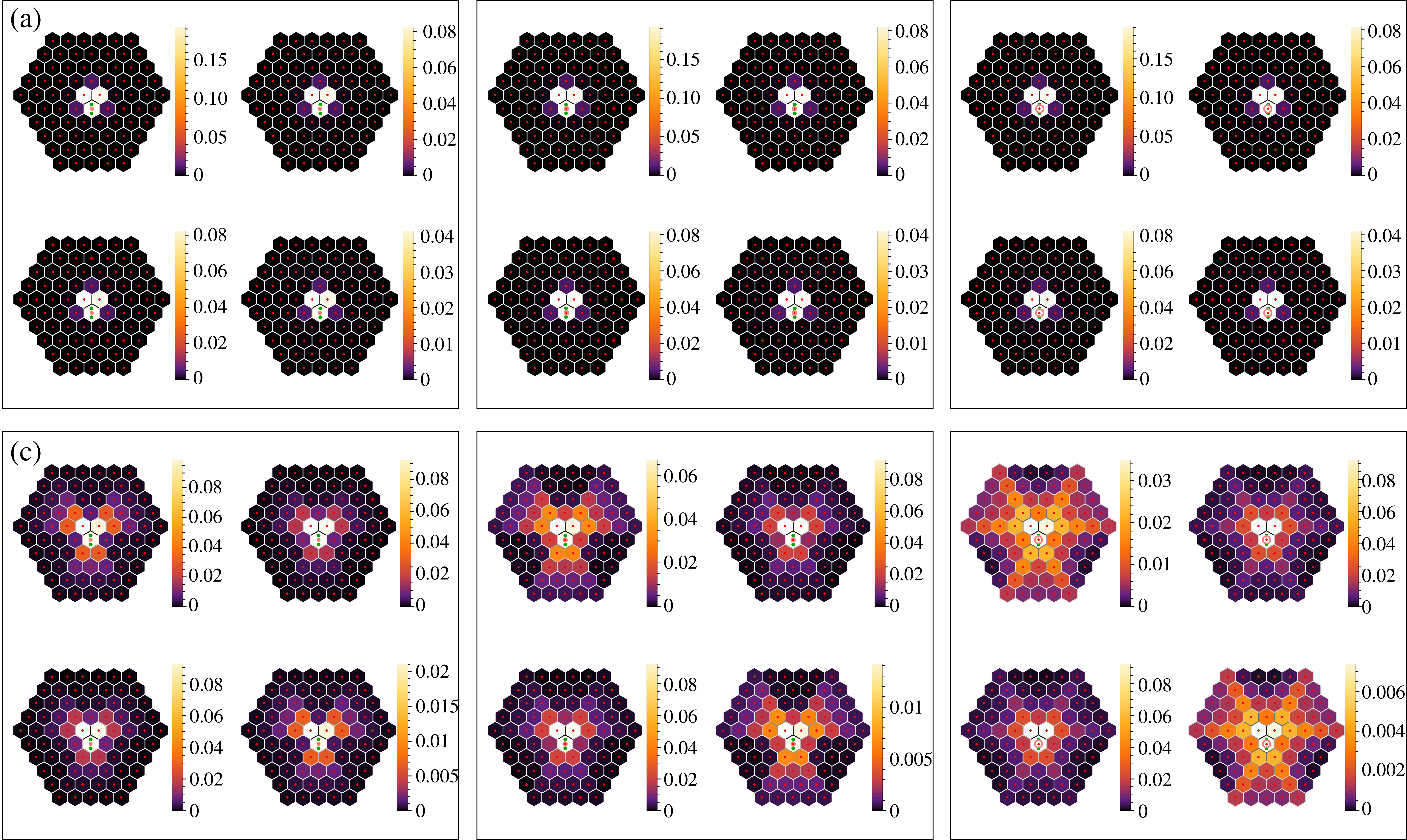}
  		\caption{Comparison of continuum projections for (a,d) $\theta=1.05^\circ$; (b,e) $\theta=0.67^\circ$, and (c,f) $\theta=0.33^\circ$. (a,b,c): hexagonal lattice; (d,e,f): relaxed lattice. Whereas the reciprocal space interlayer coupling do not depend on twist angle for the rigid lattice, relaxation gives a very long-range in the $AA$ and $BB$ couplings.
  	}\label{fig:allpro}
  	\end{figure*}
  	
Let us take a more detailed look at the continuum model projection as a function of twist angle.
As we can see in Fig.~\ref{fig:allpro} the results from the projection are indeed very different in the $AA$ and $BB$ channel after relaxation and strongly dependent on twist angle. The reason is that while the physical size of the $AA$ aligned regions is constant, independent of twist angle, and thus we would expect the real-space potential to be independent of angle,the reciprocal lattice spacing reduces substantially as we  change the angle, and thus many more Fourier components  of smaller magnitude are needed to describe this potential. The $AB$ regions, grow, leading to almost constant coupling terms.

The model that seems to describe this behaviour is a slightly extended version of the real-space model derived in the previous section: a confined potential well, where the well  is centered on the region of $AA$ alignment, with a sharp cut-off at the edges. This is very difficult to describe in momentum space, but the real space wave functions all look very similar, independent of twist angle. We have not pursued such a model here, since we know we can do very accurate tight-binding and continuum model calculations at an an angle of, say, $1.05^\circ$. We can either turn that into a real-space Hamiltonian as above, 
which will then have spatially localised solutions, or we can solve the problem in $k$ space for such a large twist angle. The (real-space) solutions for the in-gap flat bands now longer dependent on twist angle, and we have thus solved the problem for these states for all smaller twist angles. It is thus absolutely incorrect to apply the lowest-harmonic Bistritzer-MacDonald model to such situations: if there is relaxation of the atomic 
lattice, this will fail drastically at small angles. The behaviour is more like states localised at the $AA$ impurity.

  	\section{Charge density}  
Since experimental data suggest twisted hBN is electrically charged, it would be interesting to try an understand the charge density in detail.
It appears that the only way to get reliable results is to sum over \emph{all} occupied states, since converged results are only found when finding summing over all eigenvectors. 	
  	 \begin{figure}
  		\includegraphics[width=\columnwidth]{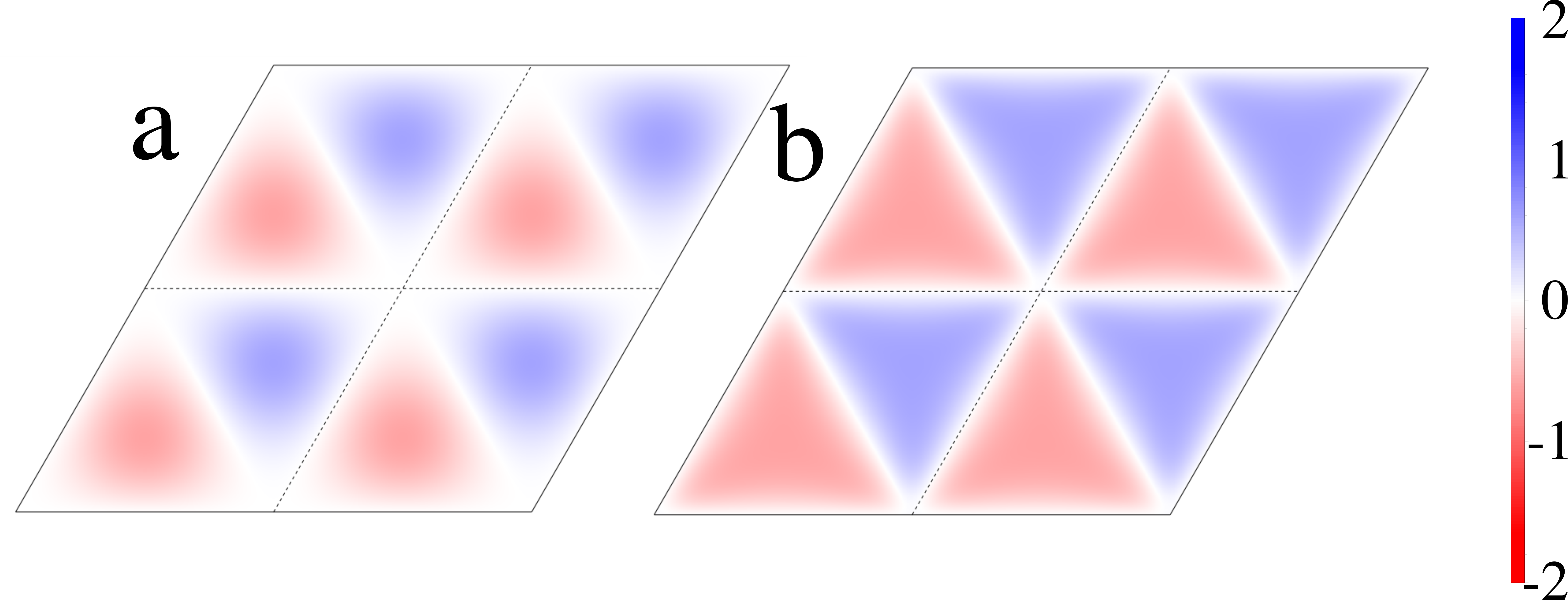}
  		\caption{Twist induced charge density at neutrality in the top layer for $\theta=1.05^\circ$. a) unstrained layer as in a) in  Fig.~\ref{fig:32x31_1}.  
  		b) relaxed layer as in c) in  Fig.~\ref{fig:32x31_1}. We clearly note the enhanced triangular symmetry.   The units used are the same as in Fig.~\ref{fig:piezo}.
  	}\label{fig:32x31_dens}
  	\end{figure}
  	
  	As can be seen in Fig.~\ref{fig:32x31_dens}, for the triangular relaxation the charge density indeed has the triangular  pattern observed in experiment. This  charge density is largely carried by the $B$ atoms, due to the difference in hopping parameters. There is no charge density for the anti-aligned twisted case, and a three orders of magnitude smaller charge density when we relax the anti-aligned crystal.
  	 
  	 \begin{figure*}
  		\includegraphics[width=0.8\textwidth]{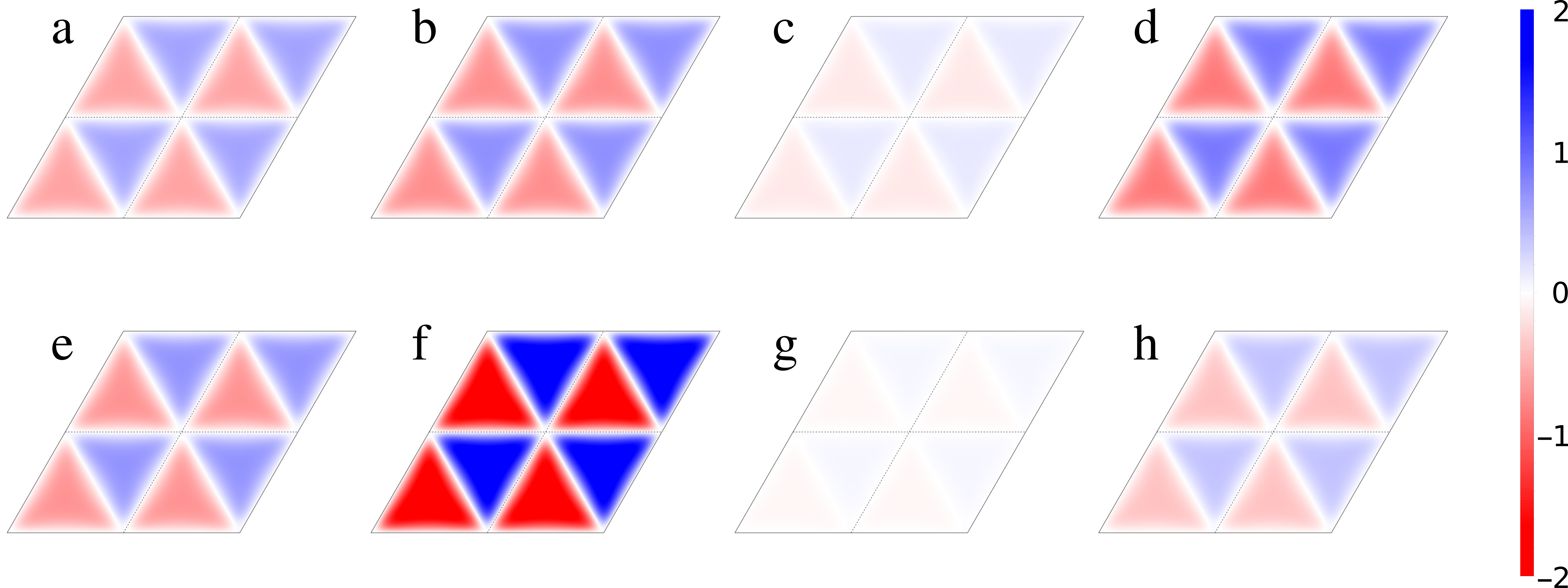}
  		\caption{Twist induced charge density at neutrality in the top layer for $\theta=1.05^\circ$ for a relaxed layer for change to the basic parameter choice $t_{BB}=0.7$, $t_{NB}=0.3$, $t_{NN}=0.15$, $\Delta=8\,\text{eV}$, $\alpha=4.4\,\text{\AA}^{-1}$ a) basic parameters
  		b) $t_{BB}=0.5\,\text{eV}$;
   		c)  $\alpha=2.2\,\text{\AA}^{-1}$;
  		d)  $\alpha=6.6\,\text{\AA}^{-1}$;
  		e) $\Delta=6\,\text{eV}$;
  		f) $t_{NB}=0.5\,\text{eV}$;
  		g) $t_{NB}=0.15\,\text{eV}$;
  		h) $t_{BB}=1.0\,\text{eV}$;   The units used are the same as in Fig.~\ref{fig:piezo}.
  	}\label{fig:32x31_params}
  	\end{figure*}

The uncertainty and sensitivity to parameters of the charge density needs  quantification, and we 
have performed a more detailed analysis, see Fig.~\ref{fig:32x31_params}.
There is some sensitivity to the range parameter $\alpha$ in \eqref{eq:hopping}: The charge density falls with an increase of the range parameter. We conclude that the experimental results suggest a relatively short-range hopping, maybe driven by many-body screening as in Ref.~\citep{GuineaWalet2019}. However, the dominant parameter is the hopping $t_{NB}$: increasing that from $0.3$ to $0.5\,\text{eV}$ increases the maximum in the electron density to $2\times 10^{12}\,\text{cm}^{-2}$. This value of the hopping is still quite reasonable, and may well help us to put constraints on microscopic calculations of such parameters.

For the case with the highest electron density,
the charge density is about $\sigma=\pm2\times 10^{12}\times10^4\times 1.6\times10^{-19}=\pm3\times 10^{-3}\mathrm{C/m}^2$.
If we take the vacuum value $\epsilon_r=1$ between the hBN layers \cite{Latini2015,Thygesen_2017}, we find, assuming the triangular domains are large enough to apply an infinite-parallel-plate approximation,  
\begin{equation}
V=\frac{\sigma}{\epsilon_r\epsilon_0} d=
\frac{3\times  10^{-3} \times 3.33 \times 10^{-10}}
{ 8.85\times 10^{-12}}=110\,\text{mV}\,.
\end{equation}

%  	\begin{figure}
%  		\includegraphics[width=\columnwidth]{chargesep.pdf}
%  		\caption{Charge density in the top layer of thBN at neutrality for $\theta=1.05^\circ$, corresponding to a) in  Fig.~\ref{fig:32x31_1}. Includes all states with $E<0$. Left: B atomic positions, right N. Units are fractional occupancy (electrons per unit cell).}\label{fig:32x31_dens2}
%  	\end{figure}

\section{Effect of charge on  electronic spectrum}
Of course such a charge density can impact the electronic spectrum of the in-gap states especially since the Wannier states are localised at the point where the positive charge density meets the negative one.
Clearly, in this case we need to look at both the ionic and the piezo-electric charges, since they are both of similar magnitude near the AA/AA' points.
The total charge-carrier density does not exceed the value of $n_0= 2\times 10^{16} \text{cm}^{2}$. Define a dimensional carrier density $\bar{n}(\vec x)=n(\vec x)/n_0$.
Expressing all distances in Angstrom, we find that the Coulomb force due to the charge in the two layers is
\begin{align}
V(\vec x)&=
\frac{\alpha}{\epsilon_r} n_0\hbar c \int \bar n(\vec y) 
\left(\frac{1}{|\vec x -\vec y|}-\frac{1}{|\vec x -\vec y+d\vec e_z|}\right)\mathrm{d}^2 y\nonumber\\
&=(3 \text{meV/\AA})\int \left(\frac{1}{|\vec x -\vec y|}-\frac{1}{|\vec x -\vec y+d\vec e_z|}\right)\mathrm{d}^2y\,.
\end{align}
Using a model for large charge domains, where just 6 domains meet at a point, we
find that the integral above is at most $2\text{\AA}$.
Since the potential has positive and negative contributions, we estimate the maximum
effect of the perturbation of the charge as a function of momentum to be much less than  $1\,\text{meV}$. This clearly does not modify the isolated bands by a significant amount.

%So let us zoom in  on an area of radius $20\,\text{\AA}$ around the AA aligned regions
%  	\begin{figure}
%  		\includegraphics[width=\columnwidth]{allch.pdf}
%  		\caption{Piezo-electric charge density near the AA and AA' aligned points.
%  		a-c: aligned, a: $theta=1.05^circ$, b: $\theta=0.67^\circ$ and c: $\theta=0.33^\circ$.
%  		d-f: antialigned, d: $theta=1.05^circ$, e: $\theta=0.67^\circ$ and f: $\theta=0.33^\circ$.
%  		}\label{fig:allch}
%  	\end{figure}
\section{Conclusions} 
We conclude that twisted hBN has in-gap flat bands. If the crystal were not to relax these would be a set of equally-spaced set of levels, similar to Landau levels. As we relax we loose some of these levels, even though they remain extremely flat (to numerical accuracy, $10^{-2}\,\text{meV}$. these can be described as a continuum model with a gap, in the Bistritzer-MacDonald mould for large twist angles. For smaller twist angles this is not the correct description, since the in-gap spectrum becomes independent of twist angle, showing these are a set of states near the AA' aligned point, which is a region that becomes independent of twist angle as that decreases.

Without lattice relaxation, parallel and antiparallel alignments have identical spectra. Since as discussed in the body of the paper the lattices relax in a very different way, the spectra also are very different, but both have flat bands.

One of the surprising features of this work is that the parallel case has a permanent
dipole polarisation, which dominates the piezo-electric charge which is caused by the strain to relaxation. This agrees with the observations in Ref.~\cite{woods2020chargepolarized,stern2020interfacial,yasuda2020stackingengineered,Zheng2020}.

	\bibliography{HBN}
	\appendix
	\section{Tight binding model expressions}
		With these additional hoppings, if we truncate the interlayer coupling to atoms either placed directly above each other and their nearest neighbours, we find  Hamiltonians that are a slight generalisation of those in Ref.~\cite{RibeiroPeres11}, see \eqref{eq:tb}
	\begin{align}	
	h_{AA}&=\left(
	\begin{array}{cccc}
	-\frac{\Delta}{2} & t g(\vec k)  & t_{NN} & t_{NB}x g(\vec k)  \\
	t g(\vec k)^* & \frac{\Delta}{2} & t_{NB}x g(\vec k) ^* & t_{BB} \\
	t_{NN} & t_{NB} xg(\vec k)  & -\frac{\Delta}{2} & t g(\vec k)  \\
	t_{NB}x g(\vec k) ^* & t_{BB} & t g(\vec k) ^* & \frac{\Delta}{2} \\
	\end{array}
	\right)\nonumber\,,\\
	h_{AA'}&=\left(
	\begin{array}{cccc}
	-\frac{\Delta}{2} & t g(\vec k)  & t_{NB} & t_{NN}x g(\vec k)  \\
	t g(\vec k)^* & \frac{\Delta}{2} & t_{BB}x g(\vec k) ^* & t_{NB} \\
	t_{NB} & t_{BB} xg(\vec k)  & \frac{\Delta}{2} & t g(\vec k)  \\
	t_{NN}x g(\vec k) ^* & t_{NB} & t g(\vec k) ^* & -\frac{\Delta}{2} \\
	\end{array}
	\right)\nonumber\,,\\
	h_{AB'}&=\left(
	\begin{array}{cccc}
	-\frac{\Delta}{2} & t g(\vec k)  & t_{NN} & t_{NB} x g(\vec k)  \\
	t g(\vec k) ^* & \frac{\Delta}{2} & t_{NB} xg(\vec k) ^* & 0 \\
	t_{NN} & t_{NB} x g(\vec k)  & -\frac{\Delta}{2} & t g(\vec k)  \\
	t_{NB} x g(\vec k) ^* & 0 & t g(\vec k) ^* & \frac{\Delta}{2} \\
	\end{array}
	\right)\nonumber\,,\\
	h_{A'B}&=\left(
	\begin{array}{cccc}
	\frac{\Delta}{2} & t g(\vec k)  & t_{BB} & t_{NB}x g(\vec k)  \\
	t g(\vec k) ^* & -\frac{\Delta}{2} & t_{NB} xg(\vec k) ^* & 0 \\
	t_{BB} & t_{NB} x g(\vec k)  & \frac{\Delta}{2} & t g(\vec k)  \\
	t_{NB} xg(\vec k) ^* & 0 & t g(\vec k) ^* & -\frac{\Delta}{2} \\
	\end{array}
	\right)\,.\label{eq:tb}
	\end{align}
	Here $\Delta$ is the gap, and $t$ the in-layer hopping; $g(\vec k)=e^{i \left(\frac{\sqrt{3}  k_{y}}{2}-\frac{ k_{x}}{2}\right)}+e^{i \left(\frac{ k_{x}}{2}+\frac{\sqrt{3}  k_{y}}{2}\right)}+1$ is the standard sum of 3 phase factors usually found in  these calculations. $x$ is the suppression factor for hopping to a next-to-nearest neighbor. Unlike in Ref.~\cite{RibeiroPeres11} we assume that these parameters are the same for all alignments.
	\section{Definition of alignment measure\label{app:align}}
	\begin{figure}[htp]
\includegraphics[width=8.5cm]{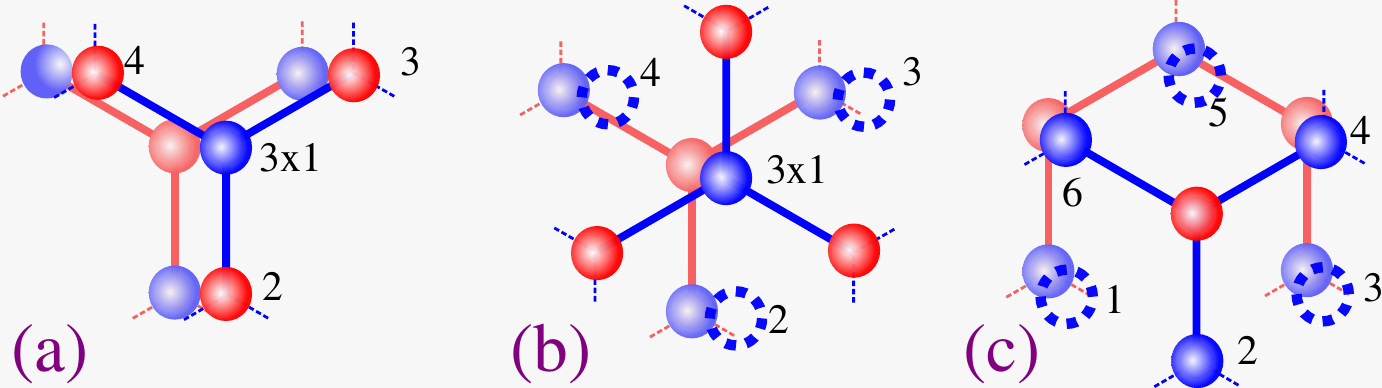}
\caption{Graphical representation of the terms used in Eqs.~(\ref{eq:wAA},\ref{eq:wAB}). The first term is for the $AA$ alignment, the last two define two situations in $AB$ alignment. The blue dotted circles are the inverted positions of the blue upper layer carbon atoms relative to the central one.}\label{fig:alignmentdef}
\end{figure}

In order to compare the size of the $AA$ and $AB$ aligned domains, we construct a measure of alignment, made of a measure for $AA$ and $AB$ alignment. 
We first define a measure of AA alignment by the function ($l$ labels the layer, $\bar{l}$ denotes the opposite layer; $\langle r_{li} r_{{\bar l}j}\rangle$ denotes the atom $j$ closest to atom $i$ but in the opposite layer; $\vec{\delta}^{(k)}_{li}$ denotes the three vectors connection atom $i$ to its nearest neighbors in the same  layer, and $\vec r_{lik\sigma}=\vec{r}_{li}+\sigma\vec{\delta}^{(k)}_{li}$)
\begin{widetext}
\begin{equation}
w_{AA}(\vec r_{li})=\frac{1}{a^2}\delta_{\langle r_{li} r_{{\bar l}j}\rangle}
\left[3\left(\vec{r}_{l,i}-\vec{r}_{\bar l,j}\right)^2+
\sum_k\left(\vec{r}_{l,i,k,+}-\vec{r}_{\bar l,j,k,+}\right)^2\right].\label{eq:wAA}
\end{equation}
In a similar way we define the quality of any AB alignment as the following function
\begin{eqnarray}
w_{AB}(\vec r_{li})&=&
\frac{1}{a^2} 
\min\left(\delta_{\langle l_i {\bar l}_j\rangle}
3\left(\vec{r}_{li}-\vec{r}_{\bar lj}\right)^2+
\sum_k\left(\vec{r}_{lik+}-\vec{r}_{\bar ljk-}\right)^2,\right.%\nonumber\\&&
\left.\sum_{k\sigma}\left(\vec{r}_{lik\sigma}-\vec{r}_{\bar{l}j}\delta_{\langle \vec{r}_{lik\sigma}, {\vec r}_{{\bar l}j}\rangle}\right)^2\label{eq:wAB}.
\right)
\end{eqnarray}
\end{widetext}
The factors of 3 in front of the central terms ensure that we use six atoms in every expression; they also weigh the central atom more heavily, which seems a sensible approach. The value of $a$ we shall use is the hBN nearest-neighbor spacing.

We then use 
\begin{equation}
w=\text{max}(w_{AA},w_{AB},w_{BA},w_{AA'},w_{AB'},w_{A'B})
\end{equation}as a measure of alignment, and we choose the colour according to the dominant choice.  
Note that $w$ is extremal for perfect alignment, negative for $AB$ and positive for $AA$ alignment.
See Fig.~\ref{fig:alignmentdef} for a graphical explanation of each of the terms.  

\end{document}